\documentclass[preprint,5p,10pt,twocolumn]{elsarticle}
\usepackage{amssymb,natbib}
\usepackage{graphicx,subfigure,amsmath,color}
\usepackage{epsfig}
\usepackage{multirow}
\usepackage{rotating}
\begin{document}
\begin{frontmatter}

\title{B$_{4/2}$ anomaly related to the shape coexistence in $^{94}$Zr}

\author[mymainaddress]{P. Buganu\corref{mycorrespondingauthor}}
\cortext[mycorrespondingauthor]{Corresponding author}
\ead{buganu@theory.nipne.ro}
\address[mymainaddress]{Department of Theoretical Physics, "Horia Hulubei"$-$National Institute for R$\&$D in Physics and Nuclear Engineering, Reactorului 30, RO-077125, POB-MG6, Bucharest$-$Magurele, Romania}
\author[mymainaddress1]{R. Benjedi}
\address[mymainaddress1]{High Energy Physics, Astrophysics and Geosciences Laboratory, Department of Physics, Faculty of Science Semlalia, Cadi Ayyad University, P. O. B. 2390, Marrakesh 40000, Morocco}

\begin{abstract}
Solutions of the Bohr-Mottelson Hamiltonian with a sixth-order anharmonic oscillator potential, for both axially-symmetric and $\gamma$-unstable deformations, are applied to investigate the lowest states of the $^{94}$Zr nucleus. The experimentally observed $B_{4/2}$ anomaly is explained, within the present approach, by the presence of a coexistence between an approximately spherical shape ($2_{1}^{+}$)  and an axially-symmetric deformed one ($4_{1}^{+}$).  On the other hand, the energies of the $0_{2}^{+}$ and $2_{2}^{+}$ states and their associated electromagnetic transitions fit better a dynamical triaxial deformation configuration ($\gamma$-unstable). This overall picture of the lowest states in $^{94}$Zr corresponds to a triple shape coexistence scenario.
\end{abstract}

\begin{keyword}
B$_{4/2}$ anomaly, shape coexistence and mixing, Bohr-Mottelson Hamiltonian, sixth-order anharmonic oscillator potential, axially-symmetric and triaxial deformations
\end{keyword}

\end{frontmatter}

\section{Introduction}
The lowest quadrupole collective states in nuclei manifest almost unique phenomena in quantum physics. For example, phases associated with the ground state deformation can be introduced, while the excited states of these nuclei are given by simple rotations and vibrations around the this equilibrium shape \cite{Casten}. Nevertheless, these are limiting cases, most of the nuclei being close to these limits or describing a shape phase transition including their critical points \cite{Cejnar}. The potentials in the deformation variables for the critical points were determined using coherent state functions on the Interacting Boson Hamiltonian \cite{Iachello1,Ginocchio,Dieperink} and further introduced in the Bohr-Mottelson Hamiltonian \cite{Bohr1,Bohr2} leading to the X(5) and E(5) critical point symmetries \cite{Iachello2,Iachello3} associated to the first and second-order shape phase transitions. Due to the flatness of the critical point potentials, the shape fluctuation phenomenon is dominant for the states of these nuclei. Beyond this phenomenon, there is a more challenging one associated with the presence of the shape coexistence and mixing \cite{Morinaga,Heyde,Garrett1,Bonatsos1}. In this case, two or more shape configurations can be found in the same nucleus. For example, the ground state band can correspond to a spherical vibrator structure, while the first $\beta$ band to an axial or a triaxial deformed configuration. Many other scenarios are possible since the phenomenon seems to be widely spread in the nuclear chart in the form of islands of shape coexistence \cite{Martinou1,Martinou2}. The shape coexistence and mixing phenomena play an important role in the structure of the states, their presence changing the pattern of the energy levels in bands as well as the strength of the inter and intra-band electromagnetic transitions compared to those of nuclei being in the limiting symmetry cases or in the transition regions. Therefore, some unusually small or large $E2$ and $E0$ electromagnetic transitions, respectively deviations from the regular patterns of the energy levels can now be well understood based on these phenomena \cite{Budaca1,Budaca2,Budaca3,Budaca4,Mennana,Benjedi1,Benjedi2}. There is also an exception, which represents one of the motivations of the present study. For example, in the last two decades, a number of nuclei have been identified manifesting the so-called $B_{4/2}$ anomaly \cite{Cakirli,Grahn,Cederwall,Goasduff,Saygi}, a situation in which the quadrupole $B(E2)$ transition between the $4_{1}^{+}$ and $2_{1}^{+}$ states is less than that between the $2_{1}^{+}$ and $0_{1}^{+}$ states. The subunitary $B_{4/2}$ ratio  of these two quadrupole electromagnetic transitions is called anomalous because none of the traditional collective models of nuclear physics anticipated such possibility. Quite recently, a progress was done in this direction by considering the Interacting Boson Model distinguishing between protons and neutrons  \cite{Teng} leaning towards triaxial deformation and band mixing as main factors for this behavior. On the other hand, a more recent study \cite{Pascu} based on a systematic done for a part of these nuclei evidenced that the triaxiality alone is not enough to get $B_{4/2}$ anomaly in all cases, while adding the condition for these nuclei to be at the border of some shape coexistence regions would increase the appearance of this phenomenon.

Since the approaches based on the Bohr-Mottelson Hamiltonian involving the sixth-order anharmonic oscillator potential proved to be appropriate to describe unusually small $B(E2)$ transitions in the ground band for highest states based on the presence of the shape coexistence with mixing, it comes naturally to apply them also for this type of nuclei manifesting the $B_{4/2}$ anomaly. Therefore, a first attempt for the present approaches is considered for the $^{94}$Zr nucleus where $B_{4/2}=0.181$. It's worth mentioning that the study of this nucleus has been considered also in relation to a recent study \cite{Marchini}, where new experimental results have been reported and investigated for the presence of the shape coexistence involving a quadrupole sum rule analysis \cite{Zielinska,Kumar}, the Monte-Carlo shell model (MCSM) \cite{Togashi} and the Interacting Boson Model-Configuration Mixing (IBM-CM) \cite{Ramos,Gavrielov}. In this latter study, the $B_{4/2}$ anomaly is not calculated within the IBM-CM, being outside the model space, but in turn is predicted an almost spherical shape for the ground state, respectively a triaxial-oblate deformation for the first excited $0_{2}^{+}$ state. These findings, as well as those established in \cite{Pascu}, are of reference for the present study where the Bohr-Mottelson Hamiltonian with a sixth-order anharmonic oscillator potential, for both symmetric-axially and $\gamma$-unstable deformations \cite{Budaca3,Budaca4}, is applied for the lowest states of the $^{94}$Zr nucleus. The goal here is to reproduce the observed $B_{4/2}$ anomaly and to establish its relation to the presence of both shape coexistence and triaxial deformation. In the present study, the triaxiality is considered through the $\gamma$-unstable approach which corresponds to a dynamical triaxial deformation.

The plan of the work is the following. After introduction, the two approaches involved in the study are briefly presented in Section 2. The Section 3 is dedicated to numerical applications for the recent experimental data of the $^{94}$Zr nucleus, respectively to the discussion of these results in relation to other studies. Finally, the Section 4 includes the main achievements of the study.

\section{Bohr-Mottelson Hamiltonian with a sixth-order anharmonic oscillator potential}

The Bohr-Mottelson Hamiltonian, written in the intrinsic reference frame, is \cite{Bohr1,Bohr2}:
\begin{eqnarray}
\hat{H}&=&-\frac{\hbar^{2}}{2B}\left(\frac{1}{\beta^{4}}\frac{\partial}{\partial\beta}\beta^{4}\frac{\partial}{\partial\beta}+\frac{1}{\beta^{2}\sin3\gamma}\frac{\partial}{\partial\gamma}\sin3\gamma\frac{\partial}{\partial\gamma}\right)\nonumber\\
&&+\frac{\hbar^{2}}{8B\beta^{2}}\sum_{k=1}^{3}\frac{\hat{L}_{k}^{2}}{\sin^{2}\left(\gamma-\frac{2}{3}\pi k\right)}+V(\beta,\gamma),
\end{eqnarray}
where $\beta$ and $\gamma$ are the intrinsic deformation coordinates, $B$ is a constant mass parameter, while $\hat{L}_{k}$ are the angular momentum projections in the intrinsic reference frame. In the present study are involved two solutions previously proposed in \cite{Budaca3,Budaca4} by considering both axially-symmetric and $\gamma$-unstable deformations. In these cases, after the separation of the variables and solving of the angular part of the eigenvalue problem, one has the following equation for the $\beta$ variable \cite{Fortunato}:
\begin{equation}
\left[-\frac{1}{\beta^{4}}\frac{\partial}{\partial\beta}\beta^{4}\frac{\partial}{\partial\beta}+\frac{\omega}{\beta^{2}}+v(\beta)\right]F(\beta)=\varepsilon F(\beta).
\label{eqbeta}
\end{equation}
Here, $\varepsilon=(2BE)/\hbar^{2}$ and $v(\beta)=(2BV)/\hbar^{2}$ are the reduced energy and potential, while with $\omega$ is denoted the constant of separation. For the $\gamma$-unstable case, $\omega=\tau(\tau+3)$ \cite{Iachello3} with $\tau=0,1,2,...$ being the quantum seniority number \cite{Bes,Rakavy}. When the axially-symmetric deformation is considered, $\omega=L(L+1)/3$ \cite{Iachello2}, where $L(L+1)$ is the eigenvalue of the total angular momentum $\hat{L}^{2}$.
In the present study, for Eq. (\ref{eqbeta}), is involved the sixth-order anharmonic oscillator potential:
\begin{equation}
v(\beta)=\beta^{2}+b_{1}\beta^{4}+b_{2}\beta^{6},
\label{pot}
\end{equation}
where, $b_1$ and $b_2$ are free parameters. Eq. (\ref{eqbeta}), for this potential, is numerically diagonalized in a basis given by solutions for an infinite square well potential \cite{Budaca1,Budaca3,Budaca4}:
\begin{equation}
f_{n\nu}(\beta)=\frac{\sqrt{2}}{\beta_{w}}\frac{\beta^{-\frac{3}{2}}J_{n\nu}\left(\frac{z_{n\nu}}{\beta_{w}}\beta\right)}{J_{\nu+1}(z_{n\nu})},
\end{equation}
where $J_{\nu}$ are Bessel functions of the first kind of index $\nu=\sqrt{\omega+9/4}$, $z_{n\nu}$ their zeros, while $\beta_w$ defines the infinite square potential \cite{Iachello2,Iachello3}. Therefore, the wave function for Eq. (\ref{eqbeta}) and potential given by Eq. (\ref{pot}) written in this basis becomes:
\begin{equation}
F_{\xi\nu}(\beta)=\sum_{n=1}^{n_{Max}}A_{\xi n}f_{n\nu}(\beta),
\label{totwav}
\end{equation}
where, $n_{Max}$ gives the dimension of the basis, $\xi=1,2,...$ quantifies the $\beta$ vibration motion, while $A_{\xi n}$ are the eigenvector components. In this conditions, the general matrix element is given by:
\begin{equation}
H_{mn}=\left(\frac{z_{n\nu}}{\beta_{w}}\right)^{2}\delta_{mn}+\beta_{w}^{2}I_{mn}^{(\nu,1)}+\sum_{i=2}^{3}b_{i-1}\beta_{w}^{2i}I_{mn}^{(\nu,i)},
\label{matrixelements}
\end{equation}
where the integrals of the potential terms are
\begin{eqnarray}
I_{mn}^{(\nu,i)}&=&\frac{2}{J_{\nu+1}(z_{m\nu})J_{\nu+1}(z_{n\nu})}\times\nonumber\\
&&\int_{0}^{1}J_{\nu}(z_{m\nu}x)J_{\nu}(z_{n\nu}x)x^{2i+1}dx,\\
&&x=\beta/\beta_{w}.
\end{eqnarray}
Having the matrix elements, one can further calculate the energy
\begin{equation}
E_{\xi,\nu}=\frac{\hbar^{2}}{2B}\varepsilon_{\xi,\nu}(b_{1},b_{2}),
\label{ener}
\end{equation}
respectively the $E2$ electromagnetic transitions using the following operator \cite{Wilets}:
\begin{eqnarray}
T_{2,\mu}^{(E2)}&=&t\beta D_{\mu,2}^{(2)}(\theta_{i})\cos\gamma+\nonumber\\
&&\frac{t\beta}{\sqrt{2}}\left[D_{\mu,2}^{(2)}(\theta_{i})+D_{\mu,-2}^{(2)}(\theta_{i})\right]\sin\gamma.
\label{tranop}
\end{eqnarray}
Here, $t=3R^{2}Ze\beta_{M}/4\pi$ depends on the nuclear radius $R=R_{0}A^{1/3}$, charge number $Z$, elementary charge $e$ and a scaling parameter $\beta_M$. Also, $R_{0}=1.2$ fm, $A$ is the atomic mass, while by $D(\theta_{i})$ are denoted the Wigner matrices \cite{Wigner}. Explicit expressions for the $B(E2)$ transitions are given in \cite{Budaca1,Budaca2}, where the two approaches involved in the present study have been proposed. The strength of these $E2$ quadrupole electromagnetic transitions are investigated in relation to the energy levels, given by Eq. (\ref{ener}), respectively to the shape of the effective potential
\begin{equation}
V_{eff}(\beta)=\frac{\hbar^{2}}{2B}\left[\frac{\omega+2}{\beta^{2}}+\beta^{2}+b_{1}\beta^{4}+b_{2}\beta^{6}\right],
\label{effpot}
\end{equation}
monopole $E0$ transition \cite{Budaca2,Budaca3,Wood}
\begin{equation}
\rho^{2}(E0;0_{2}^{+}\rightarrow0_{1}^{+})=\left(\frac{3Z}{4\pi}\right)^{2}\beta_{M}^{4}\langle F_{2\nu}(\beta)|\beta^{2}|F_{1\nu'}(\beta)\rangle^{2},
\label{monopole}
\end{equation}
and the probability density distribution of the $\beta_2$ deformation $\rho_{\xi\nu}(\beta)=|F_{\xi\nu}(\beta)|^{2}\beta^{4}$. For this latter quantity, as well as for the effective potential (\ref{effpot}), is used $\beta_{2}=\beta_{M}\beta$ in order to restore the magnitude of the quadrupole deformation in relation to the traditional microscopic domain of values.

\section{Numerical results}
The $^{94}$Zr nucleus is studied for the presence of the $B_{4/2}$ anomaly in relation to the shape coexistence phenomenon and triaxial deformation. Recently, the experimental data of $^{94}$Zr \cite{Abriola} have been revised by including new values for the $E2$ electromagnetic transitions \cite{Marchini,Zielinska2}. The strength of the electromagnetic transition is important in assigning a state to a certain band, as well as in establishing the presence of the shape coexistence and mixing. Therefore, new experimental data impose a new theoretical revision of the structure of these states. Some applications have already been made \cite{Marchini} suggesting a presence of a coexistence between an approximately spherical shape, for the ground state, and a triaxial one tending towards oblate shape for the $0_{2}^+$ state. The $^{94}$Zr nucleus presents also a $B_{4/2}$ anomaly, which was not addressed in \cite{Marchini} due to the limited model space. This latter aspect is investigated with priority here, following to verify the role played by both shape coexistence and triaxial deformation in the appearance of this anomaly. The contribution of the triaxial deformation is well-established \cite{Teng}, while new findings underline the importance of the shape coexistence \cite{Pascu}. The present approaches, proposed in \cite{Budaca3,Budaca4}, proved to be able to describe unusually small $B_{8/2}$ \cite{Mennana} and $B_{6/2}$ \cite{Benjedi1,Benjedi2}, as well as the shape coexistence phenomena for many nuclei \cite{Budaca2,Budaca3,Budaca4,Benjedi1,Benjedi2,Mennana2,Buganu,Benjedi3}. Therefore, these approaches are suitable for the proposed objective.

The parameters of the two approaches are given in Table \ref{tab1} and are set as follows. There are only two free parameters, $b_1$ and $b_2$, for each approach, which are fitted through the root mean square procedure for the reduced energies of the considered states. The reduced energy refers to the absolute energy normalized to the energy of the $2_{1}^+$ state. Therefore, the scaling parameter $\hbar^{2}/2B$ is not involved in the fit. This latter scaling parameter is fixed in the next step such that to exactly reproduce the experimental energy of the $2_{1}^+$ state. In a similar way, the scaling parameter $\beta_{M}$ is fixed to reproduce the experimental value of the $B(E2;2_{1}^{+}\rightarrow0_{1}^+)$ transition. In turn, $\beta_w$ is constrained to intersect the tail of the sixth-order potential being thus a function of the $b_1$ and $b_2$ parameters. This fitting method allows us to determine the behavior of the states only as a function of the two free parameters $b_1$ and $b_2$, while the other two scaling ones, $\hbar^{2}/2B$ and $\beta_M$, playing a role only in establishing the correspondence with the absolute values of the experimental data.

\begin{table}
\caption{Parameters of the Bohr-Mottelson Hamiltonian with sextic potential for $\gamma$-unstable and symmetric-axially deformations fitted through the root mean square procedure (rms) for the experimental data of the $^{94}$Zr nucleus \cite{Marchini,Abriola,Zielinska2}. The parameters $b_{1}$ and $b_{2}$ lead to $\beta_w$ equals 7.20 and 8.48 for the former and latter deformation, respectively. }
\begin{center}
\resizebox{8.5cm}{0.8cm} {
\begin{tabular}{cccccc}
\hline
$^{94}$Zr&\multicolumn{2}{c}{Free parameters}&&\multicolumn{2}{c}{Scaling factors}\\
\cline{2-3}\cline{5-6}
Approaches&$b_{1}$&$b_{2}$&$rms$&$\frac{\hbar^{2}}{2B}$ [MeV]&$\beta_{M}$\\
\cline{2-3}\cline{5-6}
$\gamma$-unstable&$-$0.0468&0.00067&0.104&0.593&0.049\\
axial-symmetry&$-$0.0416&0.00048&0.165&0.991&0.050\\
\hline
\end{tabular}}
\end{center}
\label{tab1}
\end{table}

\begin{figure*}
  \centering
  \includegraphics[width=1.0\textwidth]{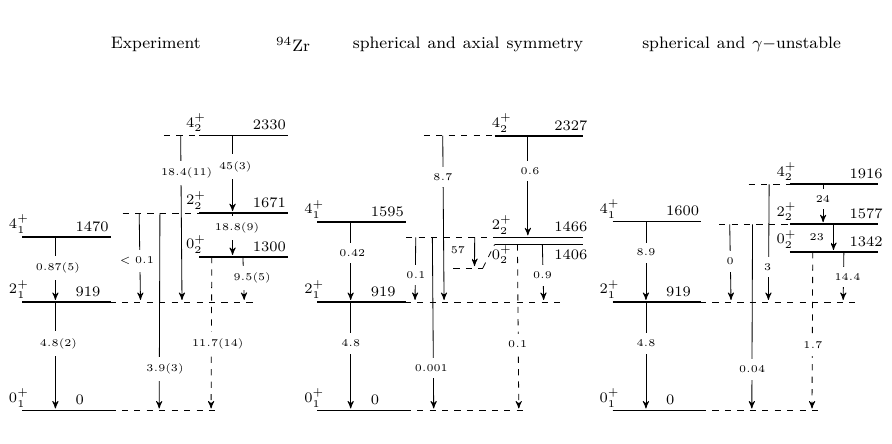}\\
  \caption{(Color online) Theoretical energy spectrum, quadrupole transitions (full arrwos) and monopole transition (dashed arrow) calculated with the two approaches are compared to the available experimental data \cite{Marchini,Abriola,Zielinska2}.}
   \label{fig1}
\end{figure*}

\begin{figure}
  \centering
  \includegraphics[width=0.45\textwidth]{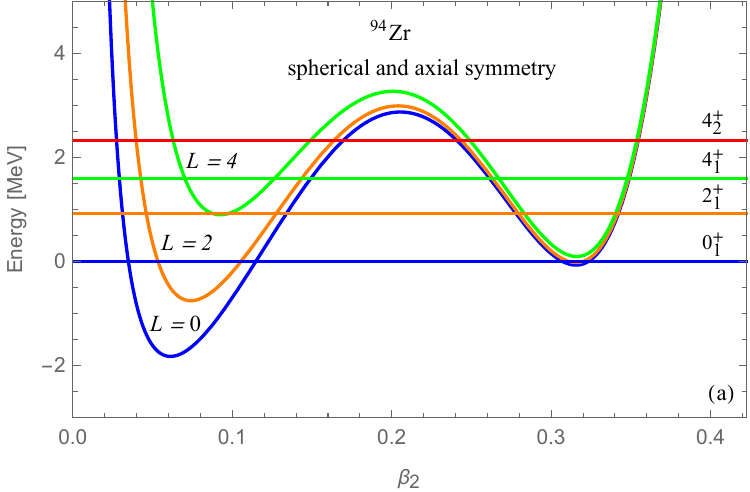}\\
  \includegraphics[width=0.45\textwidth]{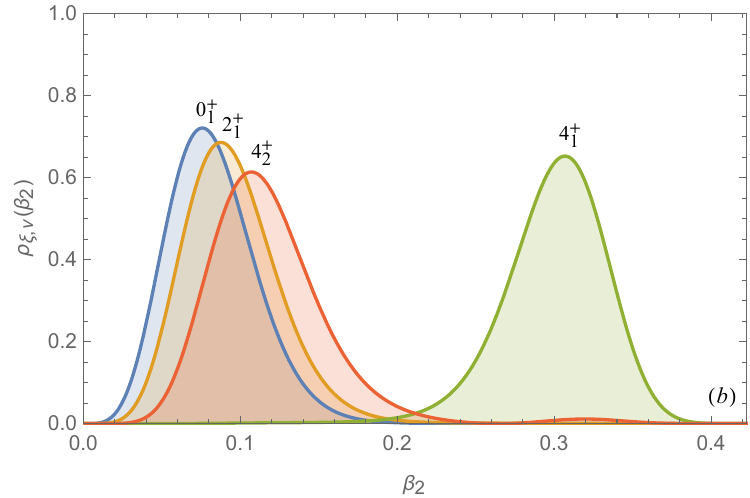}\\
  \caption{(Color online) Energy and effective potential (a), respectively  probability density distribution of deformation (b) for the $0_1^+$, $2_1^+$, $4_1^+$, $4_2^+$ states for $^{94}$Zr determined within the symmetric-axially deformed approach.}
   \label{fig2}
\end{figure}

\begin{figure}
  \centering
  \includegraphics[width=0.45\textwidth]{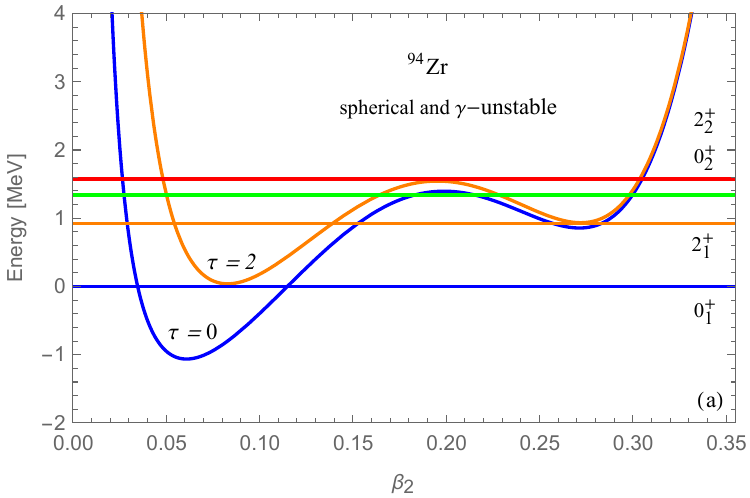}\\
  \includegraphics[width=0.45\textwidth]{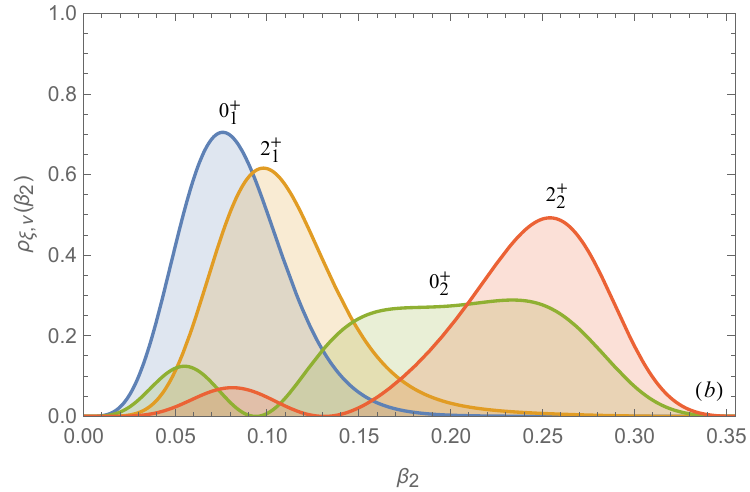}\\
  \caption{(Color online) Energy and effective potential (a), respectively  probability density distribution of deformation (b) for the $0_1^+$, $2_1^+$, $0_2^+$, $2_2^+$ states for $^{94}$Zr determined within the $\gamma$-unstable approach.}
   \label{fig3}
\end{figure}

The energy spectrum and electromagnetic transitions, calculated using the parameters from Table \ref{tab1}, are compared to the experimental data of the $^{94}$Zr nucleus \cite{Marchini,Zielinska2} in Fig. \ref{fig1}.
First, one can notice that the energies of the $4_{1}^{+}$ and $4_{2}^{+}$ states, respectively the $B(E2)$ transition between these states and the $2_{1}^{+}$ state are better reproduced by the spherical and axial symmetry approach than the spherical and $\gamma$-unstable one, while the energies of the $0_{2}^{+}$ and $2_{2}^{+}$ states and their associated $E2$ and $E0$ transitions clearly fit well the $\gamma$-unstable description. The overall picture indicates a mixing of at least two configurations of states. In order to understand better these results, one needs to correlate them with the information extracted from the plots of the effective potentials and probability density distribution of the $\beta_{2}$ deformation shown in Figs. \ref{fig2} and \ref{fig3}.
According to these figures, within both approaches the effective potential has two minima, an approximately spherical one and a well-deformed one, the latter being symmetric-axially and $\gamma$-unstable deformed, respectively. This is a clear signature for the presence of the shape coexistence between these shapes, approximately spherical, symmetric-axially and $\gamma$-unstable ones. Notable that the well-deformed minimum for the symmetric-axial deformation is deeper and consequently separated by a higher barrier from the near-spherical minimum compared the $\gamma$-unstable case. Therefore, the high barrier of the potential induces a change in the $\beta_2$ deformation for the $4_{1}^{+}$ state ($\beta_2=0.300$) of which probability density distribution of deformation has a single peak positioned above the well-deformed minimum compared to the peak of the $2_{1}^{+}$ state ($\beta_2=0.095$) which more or less preserves the ground state deformation ($\beta_2=0.083$) associated to the near-spherical minimum. With other words, according to Fig. \ref{fig2}, the $2_1^+$ and $4_1^+$ states have distinct shapes, approximately spherical and symmetric-axial ones, being separated by a high barrier of the energy potential, the latter decreasing the strength of the $B(E2;4_{1}^{+}\rightarrow2_1^+)$ transition such that the $B_{4/2}$ becomes less than unity. The calculated value of this $B(E2)$ transition is 0.42 W.u., which is very close to the experimental value of 0.87(5) W.u.. This agreement supports the above scenario, at least for $^{94}$Zr, according to which the shape coexistence can explain the presence of the so-called $B_{4/2}$ anomaly. Nevertheless, a question still remains being related to the presence of the triaxial deformation, which seems to be a clear characteristic of this category of nuclei \cite{Teng,Pascu}. Indeed, recent results \cite{Marchini} predict an approximately spherical shape for the ground state of this nucleus, which is also supported by the present results for both involved approaches. Concerning the triaxial deformation tending towards the oblate shape (symmetric axial deformation) predicted in \cite{Marchini} for the $0_2^+$ state, this is closer to the picture of the $\gamma$-unstable approach observed in Fig. 3 (b) for this states. Concretely, the probability density distribution of deformation of the $0_{2}^{+}$ state has three peaks ($\beta_2=0.193$), even if its associated wave function has one node. The two barely visible peaks, which are dominant for this state, are mainly positioned above the well-deformed minimum corresponding in this case to the $\gamma$-unstable deformation. Instead, the small peak of the $0_{2}^{+}$ state is above the near-spherical minimum being responsible for the presence of the shape mixing and consequently for the small $E0$ monopole transition represented in Fig. \ref{fig1} by a dashed arrow. This mixing of the shapes of the $0_{2}^{+}$ state explains why the monopole transition within the $\gamma$-unstable approach is closer to the experimental value than the value calculated with the symmetric-axially deformed approach. Also, this shape mixing in the $0_{2}^{+}$ state is responsible for the good reproduction of the $B(E2;0_{2}^{+}\rightarrow2_{1}^{+})$ transition, the two states having an appreciable overlapping according to Fig. 3 (b).  Concerning the $2_{2}^{+}$ state ($\beta_2=0.233$), this has a clearly dominant peak above the $\gamma$-unstable deformed minimum, which leads to a good agreement of the calculated $B(E2;2_{2}^{+}\rightarrow0_{2}^{+})$ transition with the corresponding experimental value, both involved states being well-deformed. By contrary, according to Fig. 3 (b), the overlapp of the $2_{2}^{+}$ state with the $0_{1}^{+}$ ($\beta_2=0.083$)  and $2_{1}^{+}$ ($\beta_2=0.109$) states is very small, the latter states preferring the near-spherical minimum. This leads to a very small value for the $B(E2;2_{2}^{+}\rightarrow0_{1}^{+})$ transition which overestimates the corresponding measured value. In turn, the $B(E2;2_{2}^{+}\rightarrow2_{1}^{+})$ transition is forbidden within the $\gamma$-unstable approach due to the selection rules over $\tau$ \cite{Budaca4}. Even so, the corresponding experimental value is quite small being less than 0.1 W.u.. There is another $B(E2)$ transition remained to discuss, between the $4_{2}^{+}$ and $2_{2}^{+}$ states. The difficulty in describing this transition within the present approaches consists in the fact that the energy of the $4_{2}^{+}$ state is better described by the spherical and axial symmetry approach, while for the $2_{2}^{+}$ state by the spherical and $\gamma$-unstable approach. Within the former approach, the difference in energy between these two states is larger than in the case of the spherical and $\gamma$-unstable approach. Consequently, the latter approach is closer to the experimental data regarding this transition. This $B(E2)$ transition would be better reproduced within a more general model which can treat on equal footing the near-spherical, symmetric-axially and $\gamma$-unstable (or $\gamma$-stable triaxiality) deformations. This treatment is suggested by the overall present results, of which complementary contribution clearly indicate a presence of a triple shape coexistence in the $^{94}$Zr nucleus. This type of coexistence can reproduce well the so-called $B_{4/2}$ anomaly as well as other inter and intra-band electromagnetic transitions. It is worth mentioning that the $\beta_2$ values, involved in the above discussion, are average values of this deformation calculated to show more clearly the difference in the deformation of the states. It is interesting that both approaches estimate the same $\beta_2=0.083$ for the ground state, which is very close to the semi-experimental value 0.088(10) \cite{ENSDF}. Also, for the $2_1^+$ state the two approaches predict a value of the $\beta_2$ deformation close to 0.100. A significant change in the $\beta_2$ deformation emerges for the other excited states being related to the presence of the symmetric-axially and $\gamma$-unstable deformations.

\section{Conclusions}
A new mechanism for the appearance of the $B_{4/2}$ anomaly is proposed in the present study based on both shape coexistence phenomenon and triaxial deformation, which fits well the recent experimental data measured for the $^{94}$Zr nucleus. Two approaches are involved in this study, both based on the collective Bohr-Mottelson Hamiltonian with a sixth-order anharmonic oscillator potential including symmetric-axially and $\gamma$-unstable deformations. The results indicate a shape coexistence between an approximately spherical ($2_{1}^{+}$) and a symmetric-axial ($4_{1}^{+}$)  deformations, which is behind the $B_{4/2}$ anomaly. The $4_{2}^{+}$ state is also near-spherical deformed, while the $0_{2}^{+}$ and $2_{2}^{+}$ states present a clear dominant characteristic of $\gamma$-unstable deformation with small mixing of spherical component. This triple shape coexistence seems to be essential in understanding the level structure of the states and the strength of the electromagnetic transitions connecting them in $^{94}$Zr. There are also some overestimated $B(E2)$ transitions, which indicate that a more general model is necessary for the treatment on equal footing of these three shape configurations.

\section*{Acknowledgments}
This work has been supported by a grant of the Ministry of Research, Innovation and Digitization, project number PN-23-21-01-01/2023.

\end{document}